\documentstyle[prl,aps,multicol,epsf]{revtex}

\begin{document}

\title{Quantum criticality and the metal-insulator transition in 2D:
a critical test.}

\author{B. Hosseinkhani and J. Zaanen}

\address{Instituut Lorentz for Theoretical Physics, Leiden University, POB 9506, 2300 RA Leiden, The Netherlands}

 \date{\today}
\maketitle

\begin{abstract}
Using recent insights obtained in heavy fermion physics 
on the thermodynamic singularity structure associated 
with quantum phase transitions, we present here an experimental
strategy to establish if the zero-temperature transition in
the disordered two dimensional gas is a real quantum phase transition.
We derive a overcomplete set of  scaling laws relating  the density and 
temperature dependence of the chemical potential and the effective mass,
which are in principle verifyable by  experiment.   
\end{abstract}


\begin{multicols}{2}

The observation of a metal-insulator transition (MIT) in the two
dimensional electron/hole gasses with quenched disorder 
(2DE/HG's)\cite{Kravchenko}
 revived the interest in the nature of disordered, interacting fermion 
systems\cite{Abrahams}. A consensus has emerged
that the transition is between a disordered Wigner crystal (Wigner
glass) at low electron density and a strongly interacting Fermi-liquid 
state at higher densities. However, the nature of the phase 
transition itself is subject of much debate. One viewpoint is that the
interacting electron system near the transition is characterized by
strong density inhomogeneities with puddles of high density electron
gas immersed in a low density insulating background\cite{ShiXie,Spivak0}, 
suggesting a transition of the  percolation kind\cite{DasSarmaper}. 
On the other hand, it has been suggested that the transition is a genuine 
continuous  quantum phase transition (QPT)\cite{Sachdev}, finding support in 
for instance the scaling collapse of
the resistivity\cite{Kravchenko,Abrahams}. In the past only transport
quantities were accessible by experiment, but recently it appeared possible
to also measure thermodynamic quantities: the inverse 
compressibility\cite{Ilani1,Dultz1,Ilani2,Dultz2}
 and the effective mass\cite{klapwijk0,klapwijk1},
both of which appear to diverge near the transition.

Thermodynamical properties are often easier to interpret than 
transport quantities. Very recently it was realized by
Si, Rosch and coworkers in the context of heavy fermion physics
that thermodynamics becomes extremely revealing in the vicinity 
of QPT's\cite{sirosch,grueneisen}.   
Based on almost trivial scaling considerations, they demonstrated
that an unambigous diagnostic can be constructed for a QPT, in the form
of a plethora of scaling laws involving thermodynamic properties.
The fact that thermodynamics is more informative at a 
strongly interacting QPT  than at a thermal phase transition
is caused by the peculiar role of temperature as an effective
finite size in the former. The other important quantity  is
the zero-temperature control parameter, tuning
the system through the QPT. This coupling constant determines which
thermodynamic quantities reveal the quantum singularity. As we
first realized in the context of cuprate superconductivity\cite{zahoscup}, 
the electron density is a particularly interesting control parameter,
because the singularity shows itself through
the temperature- and density  derivatives of the chemical potential,
in addition to independent information  contained in the specific heat.
It are precisely these quantities which have become accessible
by experiment in the MOSFET's and we will present here a recipy
for how to use this information to uncover the QPT, if present
in these systems. 

The thermodynamic singularity structure of a QPT is of a different kind
than  the one encountered at thermal phase transitions. The role of temperature 
is taken by the coupling constant $y$, a quantity tuning the system through
its zero temperature transition at $y_c$. We take as a working definition
for a `genuine' QPT that (a) at zero temperature the quantum dynamics
in $D = d+z$  dimensional space time($z$ is the dynamical exponent) becomes
scale invariant at $y_c$, and (b) this critical state is universal in the
sense that it obeys hyperscaling. In Euclidean path-integral representation,
temperature sets the inverse compactification radius of the imaginary
time direction and acts therefore as a finite size. These postulates can
be compactly represented in the form of a scaling relation for the singular 
part of the free energy,        
     
\begin{equation}
F_s ( r, T) = b^{- ( d + z)} F_s ( b^{y_r} r, b^z T )
\label{freenscal}
\end{equation}

where $r = (y - y_c) / y_c$,  the reduced coupling constant. 
Following Zhu {\em et al.}\cite{sirosch},  
Eq. (\ref{freenscal}) is equivalent to the following 
scaling forms for the free energy density, 

\begin{eqnarray}
F_s ( r, T) & = & - \rho_0 r^{( d + z ) / y_r} \tilde{f} \left(
 { T \over { T_0 r^{z / y_r} } } \right)  
\nonumber \\
& = & - \rho_0 \left( { T \over {T_0} } \right)^{ ( d + z ) / z}
f  \left(
 { r \over { (T/T_0)^{y_r / z} } } \right) 
\label{freeenwork}
\end{eqnarray}

where $\rho_0$ and $T_0$ are non-universal constants, while $f(x)$ and
$\tilde f (x)$ are universal scaling functions. The first form is
useful in the low temperature regime of the (dis)ordered phase in the
proximity of the QPT, while the second form describes the quantum
critical regime itself. Acoordingly, since there is no
singularity at $r = 0, T > 0$, the `quantum critical' function $f$
can be expanded
as $f (x \rightarrow 0) \simeq f (0 ) + x f' (0) + (1/2) x^2 f''(0) + \cdots$. 
On the other hand, in the stable phase the scaling function is expanded 
as $\tilde{f} (x) = \tilde{f} (0) + g ( x )$ where $g(x)$ describes the 
low temperature thermodynamics of the phases to the left- or right side 
of the QPT. By taking derivatives to the coupling constant and/or temperature
one easily derives the thermodynamic quantities. The fact that 
temperature enters through finite size scaling is a blessing; the
outcome is a highly overcomplete set of scaling laws\cite{sirosch,zahoscup}
relating the zero- and finite temperature exponents of the thermodynamic quantities
via $y_r$ and $z$.

Let us now apply this general wisdom to the 2D MIT. The coupling constant
clearly corresponds with the carrier density $n$ while the transition
occurs at a critical $n_c$ such that the reduced coupling constant
is $r = ( n - n_c ) / n_c$. The thermodynamic quantities associated
with the density are the chemical potential $\mu = \partial F / \partial n$
and the inverse compressibility (or incompressibility) $(1 / \kappa) =
(1 / n) (\partial^2 F / \partial n^2)$. In addition one has
the specific heat $C = - T (\partial^2 F / \partial T^2)$ and the 
temperature derivative of $\mu$, $\Upsilon_n = (\partial \mu / \partial T) 
= ( \partial^2 F  / \partial T \partial n)$. In principle, one can derive
a total of 9 scaling forms for the singular parts of $1/\kappa$, $\eta$
and $C$, applicable to the quantum critical regime ($r=0$, finite $T$)
and the low temperature regimes ($T \rightarrow 0$, $r \neq 0$) of the
two phases in the proximity of the QPT, respectively. Although the
physics is in any other regard very different, the 2d MIT shares 
the density as coupling constant with the (suspected) QPT in optimally 
doped cuprates\cite{cuprcrit}
and one can find a full set of scaling forms in a paper\cite{zahoscup}
dealing with the specifics of the cuprates.

Another difference with the cuprates
is that the quantities of crucial importance (compressibility,
chemical potential) are now routinely measured in the 
MOSFET's\cite{Ilani1,Dultz1,Ilani2,Dultz2}  while
this information is not available yet in the cuprates\cite{zahoscup}.
In the remainder we will be primairely  interested in the 
high density side of the MIT. All we have to know is the function
$g(x)$ in this regime, and this can be deduced from the low temperature
behavior of the specific heat $C$. Because of their `lack of volume' $C$
cannot be measured directly and we will assume that this state is 
`like a Fermi-liquid' in the sense that the following Fermi-liquid
relation is still valid: $(C / C_0)  =  1 + F^s_1 /3   =  (m^* /  m)$,
where $C_0 = \nu T$, the specific heat in
the non-interacting limit ($\nu$ is the density of states) while $F^s_1$
is the Landau parameter which also governs the transport effective mass
$m^*/m$. This effective mass is directly measured  in the 
magneto-transport experiments\cite{klapwijk1}, and found to be finite
at zero temperature, suggesting a linear specific heat $C = \gamma T$.
It is easily checked that this implies $g(x) = c x^{y_0 +1}$ (masless)
spectrum with $y_0 = 1$.      

The scaling relations for the various quantities are easily obtained 
and let us directly specialize to the case of the `Fermi-liquid like' 
regime in the proximity of the 2D MIT ($d=2$ and $y_0 = 1$). The unknown
exponents are the dynamical exponent $z$ and the coupling constant
(electron density) exponent $y_r$. It is convenient to reparametrize
the latter in terms of an exponent having a similar status as
the specific heat exponent $\alpha$ in the case of a thermal
phase transition\cite{zahoscup},

\begin{equation}
\alpha_r = 2 - ( 2 + z)/ y_r
\label{alphar}
\end{equation}

The singular, low temperature  contributions to the various thermodynamics 
quantities in the liquid have the following form,
\begin{eqnarray}
C_{cr} (T \rightarrow 0, r) & = & \gamma_{cr} \; T \nonumber \\
\Upsilon_{r,cr} (T \rightarrow 0, r) & = & \upsilon_{cr} \; T \nonumber \\
{1 \over {\kappa_{cr} (T \rightarrow 0, r)} } & = &
K_{cr, 0} + K_{cr, T}\;  T^2   
\label{fermiexp}
\end{eqnarray}

where the various coefficients which depend critically
on $r = ( n - n_c)/n_c $ are given by,

\begin{eqnarray}
\gamma_{cr} & = & { {2 \rho_0 c} \over {T_0^2} } 
\; r^{ ( 2 - \alpha_r) ( 2 - z) /(2 +z )}
\nonumber \\
\upsilon_{cr} & = & - { {2 \rho_0 c} \over {T_0^2} } 
 \; { {( 2 - z)} \over { (2 + z) } }  
( 2 - \alpha_r) \;  r^{ ( 2 - \alpha_r) ( 2 -  z) /(2 +z) - 1} 
\nonumber \\
K_{cr,0}  & = & - \rho_0 \tilde{f} (0) ( 1 + z) ( 2 - \alpha_r) r^{ - \alpha_r}
\nonumber \\ 
K_{cr, T} & =  & -
{ {c \rho_0} \over {T_0^2} } ( 2 - z) ( ( 2 - \alpha_r) \left( {{ 2 - z} \over { 2 + z} } 
\right) - 1 ) 
\nonumber \\
&  & \times \; r^{ ( 2 - \alpha_r) ( 2 - z) /(2 + z) - 2} 
\label{fermicoef}
\end{eqnarray}

One infers that the divergence of the prefactors $\gamma, \upsilon, K_T$
of the temperature 
dependent parts of $C, \Upsilon$ and $1 / \kappa$ 
are all governed by the fundamental mass exponent defined through
$\gamma \sim m^* / m = r^{\alpha_m}$ with 

\begin{equation}
\alpha_{m} = ( 2 - \alpha_r) ( 2 - z) /( 2 + z ), 
\label{massexp}
\end{equation}

such that $\upsilon \sim 
r^{\alpha_m - 1} $ and $K_T \sim  r^{\alpha_m - 2}$. These relations
between the exponents are just a consequence of the assumption that $r$
enters the free energy in the form of a power law
and the Fermi-liquid form for $\tilde{f}$. Hence, by measuring with 
high precision the temperature dependence of the chemical potential 
and incompressibility  one can  
critically test if the system is like a Fermi-liquid with a
mass diverging at the MIT. According to the transport 
measurements\cite{klapwijk0,klapwijk1}
$m^* \sim r^{\alpha_m}$ with $\alpha_m = - 0.5 \pm 0.1$ and when the
interpretations are correct it has to be that $\upsilon \sim r^{-1.5}$
and $K_T \sim r^{-2.5}$.

In fact, the first genuine scaling laws in this liquid regime
can be deduced from universal {\em amplitude ratio's}. Usually 
amplitudes are non-universal but Zhu et al.\cite{sirosch} realized 
that in the case
of pressure driven QPT's the Grueneisen ratio becomes universal.
The analogon of the Grueneisen ratio in the case of the density driven 
transition is the ratio  of $\Upsilon$ and $C$ which is 
expressed entirely in the exponents while the non-universal
factors ($\rho_0, T_0, c$) drop out,

\begin{equation}
\Gamma_r = { {\Upsilon_{r,cr}} \over  {C_{cr}} } = 
\left( { {(2 - z)(2-\alpha_r)} \over {2+z} } \right) \;  r^{-1} 
\label{Gamma}
\end{equation}

In addition, Eq. (\ref{fermicoef}) implies a second universal ratio
which has not been identified before,

\begin{equation}
\Gamma'_r = { {T \upsilon } \over {K_{T}} } =
\left( { { 2(2-\alpha_{r}) } \over { ( 2 - \alpha_r ) ( 2 - z) - 2 -z } }
\right) \; r^{-1} 
\label{Gammaprime}
\end{equation}

It follows from the scaling analysis that the
divergence of the zero-temperature incompressibility $K_0$  is governed 
directly by the elementary `energy exponent' $\alpha_r$ -- this is no
wonder since the incompressibility has the same status as a specific
heat when the roles of temperature and density are exchanged. Hence,
the zero temperature incompressibility is the quantity which reveals
the quantum singularity most directly in case of density-driven transitions.
As an interesting aside, the screening length $s \sim 1/ \kappa$ in a 2D
charged liquid and this length is therefore governed at zero temperature
by the exponent $\alpha_r$. One could be tempted to identify the screening
length with the correlation length but this is not quite right: the
correlation length is governed instead by the exponent $\nu = 1/ y_r$.
In fact, the correlation length is the length scale associated with 
order parameter correlations and the screening length has a completely
different, `truely thermodynamic' status. There might well be an
order parameter at work at the MIT but we have left it completely
implicit in the present analysis which is equivalent to the finite size
scaling of the specific heat singularity at a thermal phase transition.      

Up to this point we have indentified three scaling laws governing 
the liquid: one relating  $\alpha_r$ to the 
mass exponent $\alpha_m$ via Eq. (\ref{massexp}), and
the universal ratio's Eq.'s (\ref{Gamma},\ref{Gammaprime}). This
is not all, because an other set of scaling relations
can be deduced governing the quantum critical regime itself.  
At the quantum-critical point the scales associated 
with the stable fixed points 
should collapse. Approaching the MIT from the metallic side the situation
is in this regard analogous to what is found in the heavy fermion systems:
the Fermi-energy (the scale of the Fermi-liquid) has to 
vanish\cite{coleman},  or equivalently, the effective mass has to diverge. 
In the quantum critical regime
the temperature dependence of the  thermodynamic properties are governed 
by a different set of scaling laws which are also governed by the  
fundamental exponents ($z, \alpha_r$).
The quantum critical regime can be accessed right at the QPT,
but also away from the critical point, as long as temperature exceeds
the Fermi-energy\cite{Sachdev}. 
Hence, if the MIT is governed by a genuine QPT it has
to be that upon approaching the phase transition a temperature  window 
opens up of universal behaviors bounded from the below by the Fermi-energy
and from the above by a cross-over to a non-universal ultraviolet.      

The temperature dependences of $C$, $\Upsilon$ and $1/\kappa$ in this
quantum critical regime are easily deduced from the second form 
for the singular part of the free-energy in Eq. (\ref{freeenwork}),

\begin{eqnarray}
C_{cr} (T, r= 0) & = & 2 \rho_0 f (0) { { (2+z) } \over {z^2} } \;
\left( {  T \over {T_0} } \right)^{ 2/z}
\nonumber \\
\Upsilon_{cr} (T, r= 0) & = & - { { \rho_0 f' (0)} \over {T_0}}
\; { {1 - \alpha_r} \over {2 - \alpha_r} }
\;   { { (2+z) } \over {z} }
\nonumber \\ 
& & \times \left( {  T \over {T_0} } \right)^{ ( 2( 1 - \alpha_r ) -z) / 
(z( 2 - \alpha_r )) } 
\nonumber \\
 { 1 \over {\kappa_{cr} (T, r=0)} }  & = & - \rho_0 f'' (0) \;
\left( {  T \over {T_0} } \right)^{ ( (2+z) \alpha_r ) / 
(z( 2 - \alpha_r )) }
\label{critscaling}
\end{eqnarray}

It follows that the temperature exponent of the specific heat directly
reveals the dynamical exponent $z$ -- a direct consequence of the role of
temperature as a finite size. Since there is no a-priori relation 
between transport and specific heat in the quantum critical regime 
 it appears that this quantity is beyond the reach of experiment.
However, again the chemical potential is a valuable source of
information. Its temperature and density derivatives should obey
two other independent scaling laws.      

In summary, we have identified five scaling relations depending
on two unknown exponents ($\alpha_r, z$), governing
the behaviors of quantities which can be measured in principle
by present day experimental techniques, which should be obeyed 
if a genuine QPT is at work in the 2DEG's. How does the above relate to 
the thermodynamic experiments, feasible in the MOSFET's? 
High precision measurements of the effective mass and
the chemical potential in the same samples are required, to
establish if these quantities reveal divergences at the same 
critical density.  The dynamical range should be improved to
densities much closer to the transition and a main message of the
present work is that the temperature dependence of especially the chemical
potential contains valuable information. 

What can be said on basis of the 
data which are at present available? The divergence of the effective
mass seems to be reasonably well established, suggesting a mass
exponent $\alpha_m = -0.5 \pm 0.1$\cite{klapwijk0,klapwijk1}. 
The incompressibility data suggest that $K_0$ diverges at some critical 
density\cite{Ilani1,Dultz1,Ilani2,Dultz2}. However, the
dynamical range covering this divergence is in the present data 
too small. Pending the choice of $n_c$ we find that the incompressibility 
upturn can equally well be fitted with an $\alpha_r$  in the range 
$0.2 - 2$\cite{thesis}. As we already emphasized, $\alpha_r$ is analogous 
to the specific
heat exponent of thermal phase transitions, and these are generically
small compared to unity. Even within the uncertainties,
the present data seem to suggest that the
incompressibility divergence is 
much stronger than the specific heat divergence at a typical classical
phase transition. It is easy to derive some bounds on the allowed
values of the exponents -- if these are violated one is surely not
dealing with a QPT. The fundamental exponent is the coupling constant
dimension $y_r = ( d + z) / ( 2 - \alpha_r)$. To ensure the relevancy
of the coupling constant $y_r > 0$ and this implies an upper bound
$\alpha_r < 2$. 

Another bound follows from
the condition that both the effective mass and the incompressibility
are diverging at the transition.
 The divergence of the mass implies $\alpha_m = -| \alpha_m |$, while a
diverging incompressibility needs  $ \alpha_r > 0$.      
Using the scaling law Eq. (\ref{massexp}) to express $\alpha_r$ in terms
of $\alpha_m$ as $\alpha_r = 2 +  | \alpha_m | ( 2 + z) / ( 2 - z)$,
we find a lower bound on $z > ( 2 + | \alpha_m | ) / ( 1 - |\alpha_m| / 2)$. 
 For instance, taking the experimental value $|\alpha_m |
\simeq 1/2$, it follows that $z > 10/3$. This is a surprisingly large
lower bound, once again
demonstrating  that quantum phase transitions involving
fermions have little in common with their bosonic- or classical 
sibblings\cite{coleman}.

Summarizing, we have presented in the above a powerful diagnostic
to establish whether the metal-insulator transition in 2DEG's is a genuine 
quantum phase transition. We are not advocating the view that {\em it has
to be} a quantum phase transition. The only way to find it out is
by more and better experiments.  Alternatively, one could take the
viewpoint that the MIT cannot possibly be a QPT given the intrinsic
tendencies of the system to micro-phase separate. The various  arguments 
\cite{ShiXie,Spivak0} in support of this viewpoint are persuasive, 
but only so to the extent that the starting assumption is satisfied:
a good metal separated from an insulating crystal by a first order
transition which is in turn `masked' by electrostatic effects and/or the
disorder potential. In this context, the `quality' of the metal refers
to its screening length being small. In 2D, the screening length $s
\sim 1 / \kappa$ and according to a Fermi-liquid relation 
$\nu / \kappa = ( m / m^*) ( 1 + F^s_0 )$. Even when
$F^s_0$ can be ignored, the incompressibility and thereby the screening
length has to diverge in the approach to the transition,
 because the effective mass 
is divergent. It is hard to imagine how a mass divergence can arise from
inhomogeneities -- it is most likely associated with the physics of
the uniform liquid\cite{dassarmamass}. Stronger, as is obvious from the
scaling analysis, the screening length and the mass are in principle
governed by different exponents, which implies that also $F^s_0$ might
behave singularly. Microscopically this might have to do with the effect
discovered by Das Sarma in the 1980's that quenched disorder enhances 
the screening length\cite{dassarmascr}, a physics which was recently
discussed in the context of the MIT by Si and Varma\cite{SiVarma}.
The bottomline is that the homogeneous 
liquid has an intrinsic tendency to become increasingly incompressible
approaching the MIT, and thereby increasingly similar to the Wigner glass
in its electrostatic responses. This in turn will tend to diminish the 
tendency 
towards the formation of inhomogeneities, potentially paving the road for
a continuous quantum phase transition. In an upcoming publication we will
address these matters on a more quantitative level.

{\em Acknowledgements}
We acknowledge fruitful discussions with T. Klapwijk, S. Das Sarma,
C.M. Varma, Q. Si and J.M.J. van Leeuwen.

\end{multicols}

\begin{thebibliography}{9}

\bibitem{Kravchenko} S.V. Kravchenko {\em et al.},
Phys. Rev. B {\bf 51}, 7038 (1995); Phys. Rev. Lett.
{\bf 77}, 4938 (1996).

\bibitem{Abrahams} E. Abrahams, S.V. Kravchenko and M.P. Sarachik,
Rev. Mod. Phys. {\bf 73}, 251 (2001).


\bibitem{ShiXie} S. He and X.C. Xie,  Phys. Rev. Lett.
{\bf 80}, 3324 (2001); J. Shi and X.C. Xie,  Phys. Rev. Lett.
{\bf 88}, 4951 (2002).

\bibitem{Spivak0} B. Spivak, Phys. Rev. B {\bf 67}, 125205 (2003);
B. Spivak and S.  Kivelson, cond-mat/0310712; cond-mat/xxxxxx.

\bibitem{DasSarmaper} S. Das Sarma {\em et al.}, cond-mat/0406655

\bibitem{Sachdev} S. Sachdev, {\em Quantum Phase transitions}
(Cambridge Univ. Press, 1999).


\bibitem{Ilani1} S. Ilani, A. Yacoby, D. Mahalu and H. Shtrikman, 
Phys. Rev. Lett. {\bf 84}, 3133 (2000).

\bibitem{Dultz1} S.H. Dultz and H.W. Jiang,  Phys. Rev. Lett.
{\bf 84}, 4689 (2000).

\bibitem{Ilani2} S. Ilani,  A. Yacoby, D. Mahalu and H. Shtrikman, 
Science {\bf 292}, 1354 (2001).

\bibitem{Dultz2} S.H. Dultz, B. Alavi  and H.W. Jiang, cond-mat/0210584.

\bibitem{klapwijk0} A.A. Shashkin, S.V. Kravchenko, V.T. Dolgopolov,
and T.M. Klapwijk, Phys. Rev. B {\bf 66}, 073303 (2002).

\bibitem{klapwijk1} A.A. Shashkin, M. Rahimi, S. Anissimova, S.V.
Kravchenko, V.T. Dolgopolov, and T.M. Klapwijk, Phys. Rev. Lett.
{\bf 91}, 046403 (2003); A.A. Shashkin, S.V. Kravchenko, V.T. Dolgopolov,
and T.M. Klapwijk, J. Phys. A: Math. Gen.  {\bf 36}, 9237 (2003).
 
\bibitem{sirosch} L. Zhu, M. Garst, A. Rosch and Q. Si, 
Phys. Rev. Lett. {\bf 91}, 066404 (2003).

\bibitem{grueneisen} R. Kuchler {\em et al.},
Phys. Rev. Lett. {\bf 91}, 066405 (2003).

\bibitem{zahoscup} J. Zaanen and B. Hosseinkhani, Phys. Rev. B,
in press (cond-mat/0403345).

\bibitem{cuprcrit}  A.V. Chubukov and S. Sachdev, Phys. Rev. Lett. {\bf 71},
169 (1993); C. Castellani, C. di Castro and m. Grilli, 
Phys. Rev. Lett. {\bf 75}, 4650 (1995); R.B. Laughlin, Adv. Phys. {\bf 47}, 
943 (1998); C.M. Varma, Phys. Rev. Lett. {\bf 83}, 3538 (1999).

\bibitem{coleman} J. Custers {\em et al.}, Nature {\bf 424}, 524 (2003);
 P. Coleman {\em et al.}, J. Phys.: cond. mat. {\bf 13}, R723 (2001).

\bibitem{thesis} B. Hosseinkhani, {\em On the thermodynamics at the 
quantum phase transitions in two dimensional electron systems},
PhD thesis, Leiden University (2004).

\bibitem{dassarmamass} It was recently argued that this mass divergence
might even reside in the clean limit: Y. Zhang and S. Das Sarma, 
cond-mat/0312565.

\bibitem{dassarmascr} S. Das Sarma, Phys. Rev. Lett. {\bf 50}, 211 (1983).

\bibitem{SiVarma} Q. Si and C.M. Varma, Phys. Rev. Lett.
{\bf 81}, 4951 (1998).



\end{thebibliography}
\end{document}